\documentclass[a4paper,fleqn,usenatbib]{mnras}

\usepackage{newtxtext,newtxmath}
\usepackage[T1]{fontenc}
\usepackage{ae,aecompl}

\usepackage{graphicx}	% Including figure files
\usepackage{amsmath}	% Advanced maths commands
\usepackage{amssymb}	% Extra maths symbols
\usepackage{supertabular}
\usepackage{pdflscape}
\usepackage{ulem}
\usepackage{array}
\usepackage[dvipsnames]{xcolor}
\newcommand{\lta}{\mathrel{\hbox{\raise 0.4 ex \hbox{$<$}\kern
                   -1.5 ex\lower .4 ex\hbox{$\sim$}}}}
\newcommand{\gta}{\mathrel{\hbox{\raise 0.4 ex \hbox{$>$}\kern
                   -1.5 ex\lower .4 ex\hbox{$\sim$}}}}

\title[Statistical analysis of CP stars]{New catalogue of Chemically Peculiar stars, and Statistical Analysis}

\author[S.~Ghazaryan et al.]{
S.~Ghazaryan,$^{1}$\thanks{E-mail: satenikghazarjan@yahoo.de}
G.~Alecian$^{2}$
and A.~A.~Hakobyan$^{1}$
\\
$^{1}$Byurakan Astrophysical Observatory, 0213 Byurakan, Aragatsotn province, Armenia\\
$^{2}$LUTH, CNRS, Observatoire de Paris, PSL University, Universit{\'e} Paris Diderot, 5 Place Jules Janssen, 92190 Meudon, France\\
}

\date{Accepted  2018 July 12. Received 2018 July 12; in original form 2018 June 4}

\pubyear{2018}

\begin{document}
\label{firstpage}
\pagerange{\pageref{firstpage}--\pageref{lastpage}}
\maketitle

\begin{abstract}
In this paper we present a new catalogue of Chemically Peculiar (CP) stars obtained by compiling publications in which abundances of metals are provided. Our catalogue includes 428 stars for which the data were obtained through spectroscopic observations. Most of them (416) are AmFm, HgMn and ApBp stars. We have used this compilation to proceed to a statistical overview of the abundance anomalies versus the physical parameters of the stars. The Spearman's rank correlation test has been applied, and a significant number of correlations of abundance peculiarities with respect to effective temperature, surface gravity and rotation velocity  have been found. Four interesting cases are discussed in details: the Mn peculiarities in HgMn stars, the Ca correlation with respect to effective temperature in AmFm stars, the case of helium and iron in ApBp stars. Furthermore, we checked for ApBp stars using Anderson-Darling test wether the belonging to a multiple system is a determinant parameter or not for abundance peculiarities.
\end{abstract}


%%%%%%%%%%%%%%%%%%%%%%%%%%%%%%%%%%%%%%%%%%%%%%%%%%

%%%%%%%%%%%%%%%%% BODY OF PAPER %%%%%%%%%%%%%%%%%%

\section{Introduction}

The first goal of this work is to create a catalogue of Chemically Peculiar (CP) stars by compiling published abundances deduced from high resolution spectroscopic observations (for most of them) realized during last decades. This catalog is an extension of the one published by \citet[][hereafter referred to as Paper I]{GhazaryanGhAl2016} for HgMn stars, to other groups of CP stars. The second goal is to use this catalog for some statistical studies of these stars. In our knowledge, the only existing general catalogue containing information about CP stars, is the one published by \citet{2009RensonandMJ} (a revised and extended version of their previous catalogue), where a large number of known or suspected CP stars are identified, but their catologue provides only photometric data, and never abundance peculiarities. However, characterization and modeling of CP stars require the knowledge of the detailed abundance peculiarities, which can be carried out only from spectroscopic observations. Such studies are dispersed over a large number of publications and over a long period of time. A new catalogue of CP stars gathering all these results will help in further theoretical studies and modeling on those type of stars whose the peculiarities are understood in the framework of atomic diffusion processes \citep[see the book of][]{MichaudMiAlRi2015}.

CP stars, which are main sequence stars, consist in different groups of stars that are, according to the Preston's classification \citep{1974Preston}: AmFm (CP1), ApBp (CP2), HgMn (CP3), He-weak (CP4) and He-rich stars (we do not consider here the $\lambda{Boo}$ stars). All these groups have different physical properties and chemical abundances. CP stars groups are defined according to their abundance anomalies and effective temperature $T_{\mathrm{eff}}$, they are all slow rotators. In this paper we explore only the 3 groups CP1, CP2 and CP3.

AmFm stars are non-magnetic CP stars, slow rotators and with 7000K$\,\lta T_{\mathrm{eff}} \lta$\,10000K. AmFm stars are mostly binaries and characterized by underabundances of Ca, and Sc (sometimes stronger underabundance than the one of Ca) in their atmospheres \citep[see][for example]{KunzliKuNo1998,GebranGeMoRi2008}. 

HgMn stars are considered to be among the quietest CP stars: no (or undetected) magnetic fields, no mixing in their atmospheres. The existence of magnetic fields in their atmospheres is still subject to debate. Their effective temperature is in the range of about 10000K to 16000K. These stars show large overabundances of iron peak elements, which may be larger by more than 2 orders of magnitude their solar abundances. HgMn stars show systematic strong overabundances of Mn and/or Hg (this last one may be up to $10^6$ times the solar value). Rare-earth elements are also overabundant in their atmospheres. This group (CP3) is believed to be the continuation toward higher effective temperature of the AmFm sequence \citep[see][etc]{CatanzaroCaLeLe2003,DolkDoWaHu2003,AlecianAlGeAuetal2009}. 

ApBp stars have generally strong magnetic fields, with 7000K$\,\lta T_{\mathrm{eff}} \lta$\,16000K. Observed magnetic fields may be up to several tens of kG at their surface and are considered to be fossil fields \citep{AlecianAlTkNeetal2016}. ApBp stars are characterized by strong overabundances of iron-peak elements, like HgMn stars, and with very strong overabundances of heavy and rare-earth elements (up to $10^5$ times solar abundances). There are at least two subgroups of magnetic ApBp stars: the cool ones ($T_{\mathrm{eff}} \lta$\,10000K) which show rather overabundances of rare earth elements and sometime short period oscillations (roAp stars), and hot ones showing overabundances of Si \citep[see][etc]{LeckroneLe1981, RyabchikovaRyPiSaetal1999,KochukhovKoTsRyetal2006}.

In Sec.~\ref{sec:sample}, we describe our new catalogue. In Sec.~\ref{sec:comp} we compare and discuss the chemical abundances and physical parameters we have compiled, to those of the previous compilation did by \citet[][]{SmithSm1996}. We discuss also some statistical results obtained by different statistical tests for the chemical abundances in function of the physical parameters of the stars. Few noteworthy cases (for Mn, He, Ca, Fe) are discussed in Sec.\,\ref{sec:noteworthy}.

\section{The new catalog of CP stars}
\label{sec:sample}

Our catalog  of CP stars consists of 428 stars, for which abundances were determined by various authors, in most cases through high resolution spectroscopy. The CP-types are those proposed by these authors. In this catalog, 108 are identified as HgMn stars (see Paper I), 188 are identified as ApBp stars, 120 as AmFm stars. The spectral type of 11 stars is not well defined even if they are considered as CP (we marked them as uncertain), and one is a well-known Horizontal-Branch star (Feige 86). All CP stars considered in this paper are listed in the file "Table\_CPstarsList.pdf" available online. Their physical parameters, including effective temperature, surface gravity, rotation velocity, and bibliographic references are given in that file.

Because of the source of the compiled data are inhomogeneous, we tried to include as much as possible the error-bars of abundance measurements, as given by the authors. If for a given element the abundances of different ions were given, we took the mean of them for the abundance value, and the error-bars were recalculated by the mean square of the errors as in Paper I. It is important to mention that the error values given in these publications correspond usually to the internal error of the determination method use by the authors, they are not the abundance uncertainty. The real abundance uncertainty should be larger than the error bar we have collected, mainly because the abundances are generally estimated using homogeneous atmospheres, even for the most recent publications using modern techniques. They do not consider abundance stratifications that are produced by atomic diffusion, and detected in many CP stars \citep[see][among the most recent ones]{CatanzaroCaGiLeetal2016,NdiayeNdLeKh2018}. If for a given element different abundances by several authors were given we took the value from the publication where  many other elements' abundances were given to have more homogeneous dataset. The detailed abundances (element per element) for each star are provided as online data\footnote{These data, and some more may be also found on the following website http://gradsvp.obspm.fr/CPstars/CPstars\_home.html}.

In the file "Table\_CPstarsList.pdf", we provide also "multiplicity" information (binarity, belonging to a cluster, etc.) of these stars according to Simbad\footnote{http://simbad.u-strasbg.fr/simbad/} archive. In Sec.~\ref{sec:stat} we look for correlations between "multiplicity" (only for single, close binaries) and abundance anomalies.

\section{Comparison with the previous compilation}
\label{sec:comp}

\begin{figure*}
\centering
\includegraphics[width=18cm]{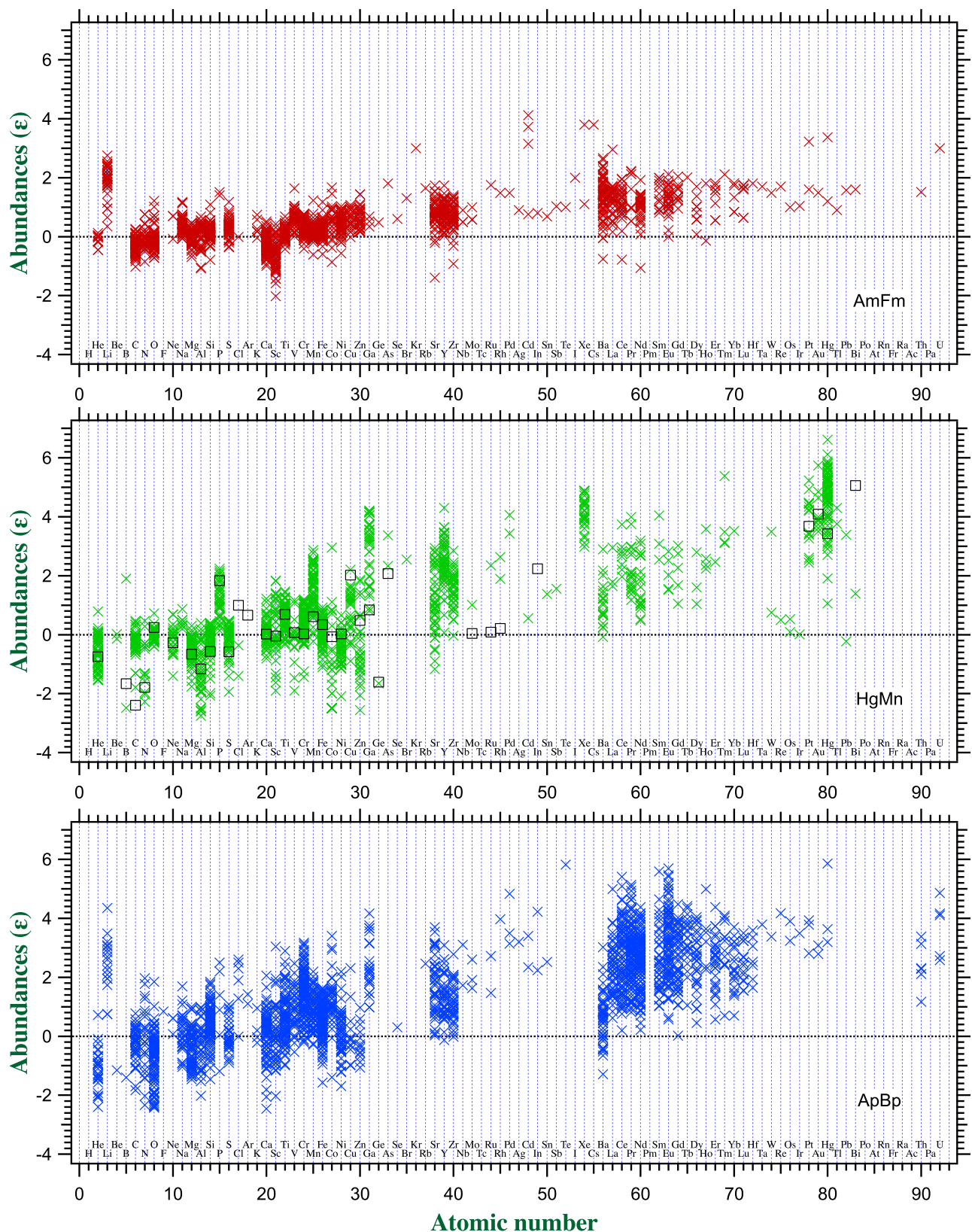}
\caption{Abundances for the three CP-Types in the present compilation vs. atomic number. Abundances ($\epsilon$) are the logarithm of the abundances divided by the solar AGS09 ones, the zero line corresponds to solar abundances. In the HgMn panel, the squares are abundances of the Horizontal-Branch star Feige 86.}
\label{fig:Layout_panels}
\end{figure*}

We compared chemical abundances of our new catalogue with the previous compilation done by \citet{SmithSm1996a} more than twenty years ago. First, to reduce the impact of the inhomogeneities between authors, abundances  were rescaled using the solar values of \citet[][hereafter AGS09]{AsplundMaGrSaSc2009}. In Fig.~\ref{fig:Layout_panels} we give abundances in logarithmic scale of $N(A)/N(H)$ (star) vs. AGS09 (Sun). Some authors give $N(A)/N_{tot}$, in these cases we have converted to the ratio $N(A)/N(H)$ assuming AGS09 solar composition for He (Y) and metals (Z), by adding 0.0357 to $\log{N(A)/N_{tot}}$.

Comparison results shows that helium [Atomic number = 2] underabundances are very scattered in HgMn and ApBp stars (by about 3 dex), and very few of them are known in AmFm stars. Note that He lines are difficult to measure for stars with $T_{\mathrm{eff}} \lta$\,10000K (only 14 of them are in our database). 

Lithium [3] appears systematically overabundant with respect to the Sun in all CP stars shown in Fig.~\ref{fig:Layout_panels}. Li is a particular element, since it is a primordial element that is easily destroyed and not produced in stars \citep[see for instance][]{SpiteSpSp1982}, except perhaps during their formation. It is destroyed in layers with temperature larger than about 2.5x$10^6$K, which happens  when Li is transported by convection from the outer layers to deep ones. This precisely occurred in the Sun, for which the superficial Li abundance is about 100 time smaller than its initial value. The superficial convection zone is due to ionization of H and He, hotter is a star smaller is this convection zone. Since H and He are already ionized in superficial layers of hot stars, the instability cannot develops. Because CP stars are hotter than the Sun, their superficial convection zone is smaller (or absent) and so, cannot transport matter deep enough to make the Li destroyed \citep[see for instance, discussion in][]{ProffittPrMi1991}. Therefore, the apparent overabundance of Li in CP stars, actually reveals the Li depletion in the Sun. The unusual overabundance of lithium was found only in Przybylsk\i's star (HD101065), which is also included in our database, but this is an exception and could perhaps be explained by magnetic field configuration in the atmospheres of that star according to \citet[][]{PolosukhinaPoShDretal2004}, or by some external process (accretion or spallation for instance).

Beryllium [4] abundance is not constrained in the atmospheres of AmFm stars, and it is marginally detected in ApBp stars. Beryllium deficiency was measured by \citep[][]{KochukhovKoTsRyetal2006} in HD133792- a weakly magnetic, evolved Ap star.  Its abundance is nearly solar in HgMn stars. For instance, its abundance was measured in the atmospheres of HD175640 and HD71066 HgMn stars by \citep[][]{CastelliCaHu2004} and \citep[][]{YuceYuCaHu2011}, respectively. 

As predicted by \citet{BorsenbergerBoMiPr1979} who estimated boron [5] stratification in the framework of atomic diffusion, this element is highly deficient in the well known HgMn star HD141556, and some others  \citep[see][for more details]{LeckroneLe1981}. However, \citet{LeckroneLe1981} observed a strong overabundance in HD78316 ($\kappa$ Cnc) which is also a well known  HgMn star, suggesting that some magnetic fields could exist in this star since atomic diffusion, which is responsible of B depletion in HgMn stars (non-magnetic stars), is strongly modified by magnetic fields.

Nitrogen [7] and oxygen [8] are less deficient respectively in HgMn and ApBp stars than in Smith's outcome. Fluorine [9] and neon [10] abundances are absent from Smith's compilation. From the literature we have found only one fluorine abundance value for the well-known 21 Com Ap star \citep[see][for more details]{MonierMoMe1990}, which is not the case for neon. It is mostly normal in HgMn stars, just a few HgMn stars shows neon underabundances in their atmospheres. Neon overabundance, which reaches up to 0.7 dex, is measured in 3 ApBp stars \citep[see][for more details]{FolsomFoWaHaetal2007,MonierMoMe1990}. Neon is overabundant in two AmFm stars - HD108642 and HD108651 \citep[see][]{KupkaKuRyBoetal1994} (the overabundance is more than 0.6 dex), and there is marginal underabundance of neon in Am star - HD198391 \citep[see][]{BudajBuIl2003}.

In the average, sodium [11] is normal in all CP stars. There is no ApBp or HgMn star with sodium abundance in Smith's compilation. Magnesium [12] is normal in AmFm stars as in Smith's paper, but it is strongly underabundant in ApBp and HgMn stars. It is overabundant in a few ApBp stars and shows large dispersion in them, which reaches to 2.5 dex. For a few HgMn stars Mg abundances wrap solar value (the abundances are +/- 0.5 around the solar value).

Aluminum [13] is absent in ApBp stars in Smith's compilation, nearly normal in AmFm stars and strongly underabundant in HgMn stars. On the contrary, in our compilation we have largely dispersed values for all CP stars. The overabundance in the ApBp stars HD95608  reaches nearly 1 dex \citep[see][]{LeBlancLeKhYaetal2015}, the underabundance in the same stars is smaller than -2 dex in HD154708 \citep[see][]{HubrigHuCaGoetal2012}. The underabundance in HgMn stars is close to -2.6 dex in HD53929 \citep[see] [for more details]{SmithSm1993}.

The underabundance of silicon [14] in HgMn stars is stronger than in the above mentioned compilation. The strongest underabundance value for HgMn stars was compiled for HD35548a from \citep{HubrigHuCaMa1999}.

Phosphorus [15] is overabundant in all CP stars, but it is absent in Smith's compilation for ApBp and AmFm stars. The highest overabundance value for ApBp stars reaches to 2.5 dex in HD166473 \citep[see][for more details]{GelbmannGe1998}. There is only one underabundant case for the Ap star HD201601 \citep[see][]{HubrigHuCaGoetal2012}.

Sulfur [16], which is absent in Smith compilation, is overabundant in AmFm stars, nearly normal in ApBp stars and deficient in HgMn stars. The strongest deficiency of sulfur was measured in HD55362 \citep[see][]{NiemczuraNiMoAe2009}, which is close to -2 dex.

Chlorine [17], argon [18] and potassium [19] are absent in Smith's compilation, but after 1993 a few abundances for these elements  were measured for the three types of CP stars. The number of measurements is so small that nothing can be said about their dispersion.

Calcium [20] dispersion is very large in ApBp stars. It is more than 4 dex in them. The highest overabundance (more than 2 dex) was measured for the ApBp star HD108945 \citep[see][]{MonierMoMe1990}. The strongest deficiency of calcium was found in HD111133, which is nearly -2.5 dex. Calcium is less underabundant in AmFm stars than in ApBp stars. A few abundances of those stars overlay solar values. Calcium is nearly normal in HgMn stars.

Scandium [21] abundances spreads around the solar value in HgMn stars, but show quite clear underabundances in AmFm stars, even more than Ca. Sc underabundances in AmFm star together with the ones of Ca is part of the clearest signatures of AmFm phenomenon. It is mostly overabundant in ApBp stars. Highest overabundance of scandium for ApBp stars was determined in HD25823, which is a bit high than 3 dex \citep[see][for more details]{BolcalBoKoDu1987}. The strongest deficiencies of scandium were found in the AmFm type star HD95608 \citep[see][]{AdelmanAdCaCaetal1999}, and in the HgMn star HD11753 \citep[see][]{MakaganiukMaKoPietal2012}. Titanium [22] dispersion in ApBp stars reaches to 4 dex, and it is generally overabundant in HgMn and ApBp stars, around solar value in AmFm stars.

The iron-peak elements (chromium [24], manganese [25], iron [26], cobalt [27], and nickel [28]) are broadly scattered slightly above the solar abundance in all the three CP-types stars we consider, sometime with significant underabundances. The strongest underabundances of chromium are determined in two HgMn stars - HD53929 and HD186122 \citep[see][for more details]{SmithSmDw1993}. In ApBp and in AmFm stars, chromium is overabundant. Manganese excess is determined in almost all CP stars, and especially in HgMn. There is a noticeable exception for the HgMn star - 36 Lyn, where Mn is deficient (see discussion in Sec.\,\ref{Mncase}). The total range in iron abundance determined in ApBp stars is more than 3 dex, which is greater than that measured for any other group of chemically peculiar stars. Iron abundance values vary by $\pm$ 1 dex around the solar value in HgMn and AmFm stars. In  most of ApBp stars Fe is overabundant with apparent correlation with the effective temperature (see discussion in Sec.\,\ref{Fecase}). Cobalt is deficient in HgMn stars, but overabundant in ApBp and AmFm stars. The strongest excess of cobalt was observed in HD116458 \citep{NishimuraNiSaKaetal2004}, which is equal to 3.4 dex. The highest deficient value of cobalt in HgMn stars was determined in two stars - HD186122 and HD174933, which is nearly -2.5 dex \citep[see][]{SmithSmDw1993}.  An unusual overabundance of cobalt was determined in the atmosphere of an HgMn star HD 143807 \citep{RyabchikovaRy1998}, which needs further detailed investigations. The HgMn stars are generally Ni-deficient although there are some examples with abundances overlaying solar values \citep{SmithSmDw1993}. On the contrary, nickel is often overabundant in AmFm stars. There is a large dispersion nearly 4 dex in ApBp stars for nickel. The highest excess of nickel was determined in HD95608 \citep{LeBlancLeKhYaetal2015} equal to 2.2 dex. It's strongest deficiency was found in HD154708 \citep{HubrigHuCaGoetal2012}, which is nearly -1.7 dex. In our compilation we have larger scatter of iron-peak elements than in Smith's compilation, which comes from the large number of observations done after 1993.

Elements heavier than iron-peak elements show increasing overabundances with atomic number for all CP stars of the three groups we consider. Copper [29], zinc [30], and gallium [31] abundances in ApBp stars were lacking in the compilation of \citet{SmithSm1996a}, we compiled a few abundances for them. These elements are overabundant in AmFm stars and in HgMn stars, except zinc, which is deficient in HgMn stars. Copper and zinc are deficient in ApBp stars, gallium is overabundant.

Germanium [32], arsenic [33], selenium [34], bromine [35], krypton [36], rubidium [37] abundances are also  absent in Smith's compilation. We have found few measurements for these elements, which is not enough to draw any conclusion for them except that they are all overabundant. 

Strontium [38], yttrium [39], zirconium [40] are generally overabundant in all CP stars except for a few of them.  The highest overabundance ($\epsilon \approx 3.7$) of strontium was found in the ApBp star HD177765 \citep{AlentievAlKoRyetal2012}, the strongest deficiency ($\epsilon \approx -1.4$) was found in the AmFm star HD225463 \citep{NiemczuraNiMuSmetal2015}. The highest overabundance of yttrium ($\epsilon \approx 4.3$) was determined in the HgMn star HD58661 \citep[see][for more details]{DworetskyDwDyPe2008}. Strong deficiency of zirconium was determined in the AmFm star HD3883 \citep{CoupryCovaBu1986}, the highest excess of it is in the HgMn star HD1909 \citep{AdelmanAdPhAd1996}.

Niobium [41], molybdenum [42], technetium [43], ruthenium [44], rhodium [45], palladium [46], silver [47], cadmium [48], indium [49], tin [50], antimony [51], tellurium [52], iodine [53], xenon [54], and cesium [55] are absent in Smith's compilation. We have found only a few overabundances of those elements (any CP type), which were determined in the two last decades. The exception is xenon, for which a lot of overabundances in HgMn stars may be found in the literature. The highest excess of xenon was determined in HD78316, strongest deficiency in HD145389 \citep[see][for more details]{DworetskyDwPePa2008}.

Concerning rare-earth elements, cerium [58], samarium [62], and europium [63] are absent in Smith's compilation for HgMn stars, and terbium [65], dysprosium [66], erbium [68], ytterbium [70], lutetium [71], hafnium [72], tantalum [73], tungsten [74], rhenium [75], and iridium [77] are absent for all CP-types. Barium [56], lanthanum [57], holmium [67], praseodymium [59], and neodymium [60] overabundances in HgMn stars in our compilation have nearly the same values as in Smith's one, but they appear larger for ApBp and AmFm stars. We have compiled a few underabundances of cerium [58], neodymium [60], and holmium [67] in AmFm stars. 

Platinum [78], gold [79], and mercury [80] overabundances are smaller in Smith's compilation than in ours. Only two osmium [76] overabundances in ApBp stars were compiled in Smith's outcome. Holmium [67], thulium [69], platinum, and mercury abundances in CP1 and CP2 stars, are absent in his compilation. Mercury abundances, which defining feature of HgMn stars, reaches up to 6.61 dex in the atmosphere of USNOA2\_0825\_03036752 \citep[see][for more details]{AlecianAlGeAuetal2009}.

We have compiled thalium [81], lead [81], and bismuth [82] overabundances in AmFm and HgMn stars, and a few excess of thorium [90] and uranium [92] in ApBp and AmFm stars, which are absent in Smith's outcome. 

This compilation confirms that the overabundances in the atmospheres of CP stars increase for heavy elements with atomic number. The large scatter of the abundances is comparable to Smith's compilation, despite of the progress in measurement techniques. As already discussed in \citet[][end of their Sec. 2]{GhazaryanGhAl2016}, and for the same reasons, we are convinced that this scatter is not only due to the heterogeneity of the data, but to the physical processes producing the abundance anomalies.

\section{Statistical analysis}
\label{sec:stat}

As we show in the previous section, abundance anomalies in the atmospheres of CP stars are often spread over by more than 1 dex for each element. This scatter may be explained by atomic diffusion and by theoretical models, it is interesting to know if they are correlated with fundamental parameters such as effective temperature, surface gravity and rotation velocity. For that reason, we applied the Spearman's rank correlation test \citep[][]{Spearman1904} between abundances and fundamental parameters\footnote{Kendall's $\tau$ test has been also applied in some cases (namely to check barely significant cases of abundance vs. gravity correlations), but we got the same results as the Spearman's one.}. The test has been applied only for each element measured in more than 11 stars (for the considered CP-type). This threshold is selected to have enough data points. In our opinion, small number statistics for other elements with less measurements make the test ineffective.

This test is widely used in astronomy. It performs a hypothesis test on a pair of two variables with null hypothesis that they are independent, and alternative hypothesis that they are not.  Usually, one accepts the alternative hypothesis of the test when p-value is less than 5 per cent (\citep[e.g.][]{FeigelsonBabu2012}. We also pay attention to those cases when the p-values are close to the adopted threshold being between 5 to 6 per cent (marginal cases). Spearman's coefficient $\rho$ is a nonparametric measure of rank correlation ($\rho\in[-1;1]$), it assesses how well the relationship between two variables can be described using a monotonic function.
When each of the variables is a perfect monotonic function of the other, the Spearman's coefficient $\rho$ is +1 or -1 \citep[see e.g.][for more details]{FeigelsonBabu2012}. Our results of the Spearman's rank correlation test  of abundances with respect to $T_{\rm{eff}}$, $\log{g}$, and $v\,\sin{i}$, are discussed in Secs.\,\ref{subsec:amfm} (AmFm), \ref{subsec:hgmn} (HgMn), and \ref{subsec:apbp} (ApBp). 

To check whether multiplicity is playing a role in the abundance anomalies, we have considered single CP stars (in this case {\it{multiplicity}} is equal to 1) and those being in binary systems (in this case {\it{multiplicity}} is equal to 2). We have applied the well-known {\it{Anderson-Darling}} (AD) test as we did in Paper I using the {\it{Kolmogorov-Smirnov}} test for HgMn stars. The null hypothesis for the two-sample nonparametric AD test corresponds to the case when two distributions are drawn from the same parent population, and the alternative hypothesis that they are not (again, with the threshold of 5 per cent for $p$-values). For more details of this statistical test, the reader is referred to \cite{EngmannCousineau11}. From this test we do not find any relation between abundance anomalies and multiplicity in HgMn stars and in AmFm stars, possibly because of the lack of data. One should mention that we checked Si, Ca, Ti, Cr, Mn, Fe, Ce, Pr, Nd, and Eu abundances dependence on multiplicity in ApBp stars (the cases with 11 and more known abundance values) and we did not find any significant result. 

\onecolumn
\begin{table}
\centering%\begin{center}
\caption{Spearman's rank test results for AmFm stars. Statistically significant correlations are shown in boldface (p-value $\le 0.05$), marginal ones are underlined ($0.05<$p-value $< 0.06$)}.
\label{table:ranksAmFm}
\begin{tabular}{|c|c|c|c|c|c|c|c|c|c|}
\hline
%&&$T_{\rm{eff}}$&&&$\log{g}$&&&$v\,\sin{i}$&\\ 
&\multicolumn{3}{c|}{$\epsilon(T_{\rm{eff}})$}&\multicolumn{3}{c|}{$\epsilon(\log{g})$}&\multicolumn{3}{c|}{$\epsilon(v\,\sin{i})$}\\ 
\hline
$Elements$&$\rho$&$p$&$N$&$\rho$&$p$&$N$&$\rho$&$p$&$N$\\
\hline
Li & \textbf{0.47} & \textbf{0.001} & \textbf{43} & -0.10 & 0.525 & 40 & 0.04 & 0.831 & 37\\
C & 0.07 & 0.524 & 76 & -0.03 & 0.804 & 76 & 0.18 & 0.132 & 69\\
N & 0.27 & 0.130 & 32 & 0.00 & 0.986 & 32 & 0.21 & 0.318 & 25\\
O & 0.15 & 0.202 & 72 & 0.01 & 0.903 & 72 & 0.11 & 0.371 & 64\\
Na & 0.10 & 0.503 & 51 & 0.26 & 0.061 & 51 & 0.07 & 0.627 & 51\\
Mg & -0.02 & 0.884 & 68 & 0.01 & 0.925 & 68 & \textbf{0.27} & \textbf{0.042} & \textbf{59}\\
Al & -0.12 & 0.385 & 52 & -0.01 & 0.922 & 51 & -0.29 & 0.070 & 39\\
Si & -0.05 & 0.635 & 84 & -0.01 & 0.895 & 82 & 0.04 & 0.715 & 71\\
S & 0.16 & 0.194 & 64 & \textbf{-0.30} & \textbf{0.019} & \textbf{62} & -0.23 & 0.096 & 54\\
Ca & \textbf{0.28} & \textbf{0.004} & \textbf{105} & 0.04 & 0.714 & 103 & 0.10 & 0.342 & 94\\ 
Sc & -0.03 & 0.844 & 64 & 0.23 & 0.069 & 64 & 0.18 & 0.194 & 56\\   
Ti & 0.20 & 0.074 & 77 & 0.20 & 0.084 & 77 & 0.21 & 0.088 & 68\\
V & -0.08 & 0.626 & 41 & -0.04 & 0.781 & 41 & 0.26 & 0.150 & 32\\
Cr & 0.10 & 0.426 & 67 & 0.13 & 0.294 & 67 & -0.22 & 0.104 & 58\\         
Mn & -0.05 & 0.696 & 60 & 0.20 & 0.133 & 60 & 0.04 & 0.785 & 52\\
Fe & -0.09 & 0.402 & 90 & 0.06 & 0.583 & 86 & -0.11 & 0.341 & 75\\ 
Co & -0.25 & 0.214 & 27 & 0.23 & 0.247 & 27 & 0.30 & 0.193 & 21\\ 
Ni & 0.08 & 0.504 & 74 & -0.08 & 0.505 & 74 & -0.11 & 0.382 & 64\\  
Cu & \textbf{0.52} & \textbf{0.009} & \textbf{24} & \textbf{0.41} & \textbf{0.049} & \textbf{24} & -0.37 & 0.072 & 24\\ 
Zn & \textbf{0.56} & \textbf{0.000} & \textbf{37} & 0.04 & 0.827 & 37 & -0.33 & 0.061 & 33\\ 
Sr & 0.03 & 0.865 & 47 & \textbf{-0.40} & \textbf{0.005} & 47 & -0.05 & 0.780 & 38\\  
Y & 0.09 & 0.460 & 65 & -0.02 & 0.905 & 65 & -0.18 & 0.182 & 57\\ 
Zr & \textbf{0.38} & \textbf{0.005} & \textbf{52} & -0.01 & 0.946 & 52 & -0.20 & 0.201 & 43\\ 
Ba & 0.06 & 0.617 & 74 & -0.08 & 0.519 & 74 & \textbf{-0.25} & \textbf{0.047} & \textbf{65}\\
La & -0.10 & 0.573 & 33 & \textbf{0.35} & \textbf{0.047} & \textbf{33} & 0.08 & 0.681 & 26\\
Ce & \textbf{0.53} & \textbf{0.001} & \textbf{36} & \underline{0.32} & \underline{0.057} & \underline{36} & -0.15 & 0.402 & 32\\
Nd & 0.28 & 0.101 & 35 & 0.24 & 0.158 & 35 & -0.06 & 0.758 & 30\\
Eu & 0.07 & 0.725 & 30 & -0.12 & 0.525 & 29 & \textbf{-0.42} & \textbf{0.041} & \textbf{24}\\ 
\hline

\end{tabular}
%\end{center}
\end{table}
%\twocolumn

%\onecolumn
\begin{table}
\centering%\begin{center}
\caption{Same as Table\,\ref{table:ranksAmFm} for HgMn stars.}
\label{table:ranksHgMn} 
\begin{tabular}{|c|*{9}{c|}}%{|c|c|c|c|c|c|c|c|c|c|}
\hline
%&&$T_{\rm{eff}}$&&&$\log{g}$&&&$v\,\sin{i}$&\\ 
&\multicolumn{3}{c|}{$\epsilon(T_{\rm{eff}})$}&\multicolumn{3}{c|}{$\epsilon(\log{g})$}&\multicolumn{3}{c|}{$\epsilon(v\,\sin{i})$}\\ 
\hline
$Elements$&$\rho$&$p$&$N$&$\rho$&$p$&$N$&$\rho$&$p$&$N$\\
\hline
He & \textbf{-0.40} & \textbf{0.001} & \textbf{65} & -0.10 & 0.450 & 65	& 0.12 & 0.355 & 64\\
C & -0.23 & 0.105 & 51 & 0.00 & 0.982 & 51 & 0.03 & 0.841 & 50\\
O & 0.23 & 0.156 & 38 & -0.28 & 0.091 & 38 & 0.15 & 0.391 & 37\\
Mg & \textbf{-0.39} & \textbf{0.001} & \textbf{69} & -0.07 & 0.580 & 69 & 0.09 & 0.489 & 68\\
Al & \textbf{-0.42} & \textbf{0.005} & \textbf{44} & 0.24 & 0.112 & 44 & \textbf{0.32} & \textbf{0.037} & \textbf{44}\\
Si & 0.01 & 0.929 & 74 & -0.15 & 0.192 & 74 & -0.01 & 0.913 & 73\\
S & \textbf{-0.61} & \textbf{0.000} & \textbf{52} & 0.05 & 0.743 & 52 & -0.03 & 0.813 & 51\\
Ti & 0.24 & 0.064 & 62 & 0.11 & 0.403 & 62 & 0.10 & 0.448 & 61\\
Cr & \textbf{-0.28} & \textbf{0.010} & \textbf{86} & -0.02 & 0.828 & 86 & \textbf{0.22} & \textbf{0.041} & \textbf{85}\\
Mn & \textbf{0.48} & \textbf{0.000} & \textbf{70} & 0.06 & 0.650 & 70 & \textbf{0.26} & \textbf{0.032} & \textbf{69}\\
Fe & 0.07 & 0.488 & 90 & 0.06 & 0.561 & 90 & -0.11 & 0.313 & 89\\
Ni & -0.05 & 0.730 & 48 & -0.01 & 0.945 & 48 & 0.11 & 0.473 & 47\\
Cu & 0.32 & 0.106 & 26 & \underline{-0.38} & \underline{0.053} & \underline{26} & 0.22 & 0.289 & 26\\
Zn & -0.18 & 0.339 & 29 & 0.00 & 0.999 & 29 & 0.12 & 0.533 & 29\\
Ga & 0.30 & 0.083 & 34 & 0.23 & 0.184 & 34 & 0.00 & 0.997 & 33\\
Sr & \textbf{-0.49} & \textbf{0.001} & \textbf{45} & \textbf{0.33} & \textbf{0.027} & \textbf{45} & 0.11 & 0.492 & 44\\
Y & -0.17 & 0.220 & 55 & 0.20 & 0.134 & 55 & 0.19 & 0.161 & 54\\
Zr & 0.25 & 0.133 & 37 & 0.25 & 0.136 & 37 & \textbf{0.38} & \textbf{0.021} & \textbf{36}\\
Xe & \textbf{0.41} & \textbf{0.032} & \textbf{28} & -0.04 & 0.823 & 28 & 0.11 & 0.568 & 28\\
Hg & \textbf{-0.22} & \textbf{0.045} & \textbf{86} & \textbf{0.27} & \textbf{0.011} & \textbf{86} & -0.01 & 0.944 & 85\\
\hline

\end{tabular}
%\end{center}
\end{table}
%\twocolumn

%\onecolumn
\begin{table}
\centering%\begin{center}
\caption{Same as Table\,\ref{table:ranksAmFm} for ApBp stars.}
\label{table:ranksApBp}
\begin{tabular}{|c|c|c|c|c|c|c|c|c|c|}
\hline
%&&$T_{\rm{eff}}$&&&$\log{g}$&&&$v\,\sin{i}$&\\ 
&\multicolumn{3}{c|}{$\epsilon(T_{\rm{eff}})$}&\multicolumn{3}{c|}{$\epsilon(\log{g})$}&\multicolumn{3}{c|}{$\epsilon(v\,\sin{i})$}\\ 
\hline
$Elements$&$\rho$&$p$&$N$&$\rho$&$p$&$N$&$\rho$&$p$&$N$\\
\hline
He & -0.32 & 0.075 & 31 & \textbf{-0.42} & \textbf{0.019} & \textbf{31} & 0.09 & 0.649 & 30\\
Li & 0.37 & 0.145 & 17 & \textbf{0.49} & \textbf{0.048} & \textbf{17} & -0.14 & 0.600 & 16\\
C & 0.24 & 0.096 & 48 & -0.04 & 0.789 & 48 & 0.21 & 0.163 & 46\\
N & 0.01 & 0.955 & 26 & -0.13 & 0.540 & 26 & -0.06 & 0.791 & 24\\
O & 0.15 & 0.108 & 109 & \textbf{0.23} & \textbf{0.017} & \textbf{109} & 0.17 & 0.148 & 75\\
Na & 0.50 & 0.004 & 31 & -0.13 & 0.499 & 31 & 0.02 & 0.903 & 30\\
Mg & \textbf{-0.35} & \textbf{0.001} & \textbf{80} & -0.14 & 0.211 & 80 & -0.15 & 0.197 & 75\\
Al & -0.20 & 0.237 & 36 & \textbf{0.38} & \textbf{0.023} & \textbf{36} & 0.05 & 0.780 & 33\\
Si & \textbf{0.49} & \textbf{0.000} & \textbf{106} & -0.17 & 0.077 & 105 & \textbf{0.27} & \textbf{0.009} & \textbf{95}\\
S & 0.23 & 0.194 & 33 & \textbf{0.36} & \textbf{0.039} & \textbf{33} & 0.34 & 0.067 & 30\\
Ca & 0.00 & 0.999 & 74 & 0.06 & 0.596 & 73 & 0.15 & 0.204 & 70\\
Ti & \textbf{0.48} & \textbf{0.000} & \textbf{88} & 0.03 & 0.758 & 88 & 0.07 & 0.555 & 83\\
V & 0.19 & 0.250 & 39 & 0.16 & 0.341 & 39 & 0.33 & 0.054 & 35\\
Cr & \textbf{0.22} & \textbf{0.016} & \textbf{120} & -0.10 & 0.257 & 119 & -0.07 & 0.443 & 113\\
Mn & \textbf{0.48} & \textbf{0.000} & \textbf{65} & -0.21 & 0.089 & 65 & 0.15 & 0.248 & 59\\
Fe & \textbf{0.53} & \textbf{0.000} & \textbf{125} & \textbf{-0.27} & \textbf{0.002} & 124 & 0.16 & 0.093 & 116\\
Co & 0.14 & 0.415 & 38 & 0.08 & 0.637 & 38 & -0.09 & 0.607 & 35\\
Ni & \textbf{0.29} & \textbf{0.030} & \textbf{56} & 0.03 & 0.828 & 55 & 0.23 & 0.102 & 53\\
Cu & 0.34 & 0.286 & 12 & 0.35 & 0.259 & 12 & 0.44 & 0.176 & 11\\
Zn & 0.27 & 0.373 & 13 & 0.09 & 0.779 & 13 & 0.08 & 0.812 & 12\\
Ga & -0.13 & 0.589 & 21 & \underline{-0.43} & \underline{0.051} & \underline{21} & -0.38 & 0.201 & 13\\
Sr & 0.13 & 0.429 & 40 & -0.13 & 0.433 & 39 & 0.00 & 0.987 & 37\\
Y & 0.14 & 0.401 & 40 & 0.03 & 0.876 & 40 & 0.31 & 0.065 & 37\\
Zr & \textbf{0.52} & \textbf{0.002} & \textbf{34} & -0.20 & 0.267 & 34 & 0.19 & 0.306 & 30\\
Ba & 0.19 & 0.168 & 52 & -0.05 & 0.746 & 52 & 0.01 & 0.939 & 50\\
La & \textbf{0.39} & \textbf{0.007} & \textbf{47} & -0.13 & 0.367 & 47 & 0.16 & 0.296 & 47\\
Ce & \textbf{0.36} & \textbf{0.001} & \textbf{77} & \textbf{0.26} & \textbf{0.022} & \textbf{77} & -0.12 & 0.301 & 74\\
Pr & \textbf{0.57} & \textbf{0.000} & \textbf{103} & 0.03 & 0.744 & 103 & \textbf{0.37} & \textbf{0.000} & \textbf{99}\\
Nd & \textbf{0.41} & \textbf{0.000} & \textbf{107} & 0.09 & 0.345 & 107 & 0.16 & 0.095 & 104\\
Sm & \textbf{0.43} & \textbf{0.007} & \textbf{38} & -0.06 & 0.737 & 38 & 0.04 & 0.837 & 36\\
Eu & \textbf{0.28} & \textbf{0.008} & \textbf{89} & \textbf{0.28} & \textbf{0.007} & \textbf{89} & -0.19 & 0.075 & 85\\
Gd & \underline{0.28} & \underline{0.055} & \underline{47} & 0.18 & 0.221 & 47 & -0.13 & 0.394 & 45\\
Tb & -0.05 & 0.829 & 22 & -0.39 & 0.071 & 22 & -0.22 & 0.323 & 22\\
Dy & 0.29 & 0.127 & 29 & -0.26 & 0.179 & 29 & -0.02 & 0.916 & 29\\
Er & 0.31 & 0.098 & 29 & -0.32 & 0.092 & 29 & 0.03 & 0.897 & 29\\
Yb & \textbf{0.72} & \textbf{0.000} & \textbf{19} & \underline{-0.44} & \underline{0.057} & \underline{19} & 0.08 & 0.739 & 19\\
\hline

\end{tabular}
%\end{center}
\end{table}  
\twocolumn

\subsection{AmFm}
\label{subsec:amfm}

Several correlations of abundance anomalies have been found for AmFm stars, they are emphasized in boldface in the Table\,\ref{table:ranksAmFm}. According to their rank, Li, Ca, Cu, Zn, Zr, and Ce abundances seems correlated with the effective temperature (see discussion about Ca in Sec.\,\ref{Cacase}). Notice that the case of Li was already pointed out by \citet{BurkhartBuCo1989} and may be understood by the destruction of this element according to the depth of the superficial convection zone (see the comment in Sec.\,\ref{sec:comp}).

Some correlations seems also to exist with respect to $\log{g}$ for \textbf{sulphur, copper, strontium, lanthanum}, and marginally for cerium, and with respect to $v\,\sin{i}$ for \textbf{magnesium, barium, europium}.

\subsection{HgMn}
\label{subsec:hgmn}

For HgMn stars, correlations with respect to $T_{\rm{eff}}$ are significant for Mg, Al, S, Cr, Mn, Sr, Xe, Hg (see Table~\ref{table:ranksHgMn}). The correlation for manganese was first noticed by \citet{SmithSmDw1993}, and we discuss it in more details in Sec.\,\ref{Mncase}. The correlation for Xe was first mentioned in Paper I, while \citet{DworetskyDwPePa2008} do not mention any significant one. On another hand,  possible correlations suggested by \citet{GhazaryanGhAl2016} for Ni, Ti and Si are not confirmed in the present study.

Considering gravity, Spearman's rank test confirms the correlation between mercury abundances and $\log{g}$ as suggested in Paper I. In addition, we find significant correlation for strontium. In the case of copper, the correlation is barely significant ($p=0.053$, underlined in Table~\ref{table:ranksHgMn}). Considering rotation velocity as parameter, correlations have been found for \textbf{aluminium, chromium, manganese, zirconium}.

\subsection{ApBp}
\label{subsec:apbp}

A large number of correlation signals emerge from Table~\ref{table:ranksApBp} for ApBp stars, as well for $T_{\rm{eff}}$ than for $\log{g}$, but specially for $T_{\rm{eff}}$. For this one, Mg, Si, Ti, Cr, Mn, Fe, Ni, Zr, La, Ce, Pr, Ne, Sm, Eu, Yb are concerned. Notice however, that the effective temperature range (7000K$\,\lta T_{\mathrm{eff}} \lta$\,16000K) for ApBp stars is much larger than for AmFm and HgMn, therefore, this may explain quite different behaviors comparing what happens at both ends. This may produce an exaggerate signal. The case of Fe is discussed in more details in Sec.\,\ref{Fecase}.

Abundances of \textbf{helium, lithium, oxygen, aluminium, sulphur, iron, cerium, europium} show a correlation with  $\log{g}$ (marginally for Ga).  The case of He is discussed in more details in \ref{Hecase}. And finally, \textbf{silicon and praseodymium} abundances seem correlated with rotation velocity.

\section{Noteworthy cases}
\label{sec:noteworthy}

Among the rather large number of correlations mentioned above, and revealed by the Spearman's test ranking, there are four noteworthy cases we would like to discuss in more details. In Fig.\,\ref{fig:Mnpec2} to \ref{fig:Fe_ApBp}, abundances are plotted vs. a stellar parameter, each point corresponds to a different star. The error bars are from the original publications (detailed in the online tables). The fitting curves are not weighted by these error bars, since most of them are missing.

\subsection{The manganese abundance peculiar feature in HgMn stars}
\label{Mncase}

\begin{figure}
\includegraphics[width=9cm]{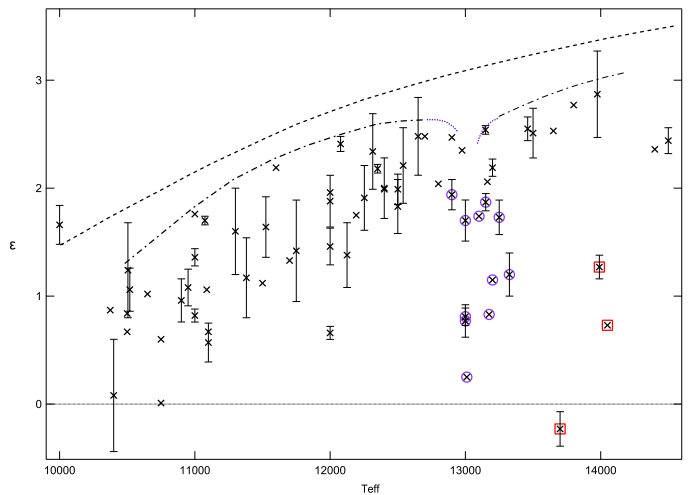}
\caption{The Mn peculiar feature. All HgMn stars of our catalog with measured Mn abundances are shown in this figure. The dashed curve is the maximum overabundance predicted theoretically by \citet{AlecianAlMi1981}. Crosses surrounded by a circle are stars that are inside what we call the group A, those surrounded by a square are the 3 hottest stars forming the group B. These highlighted stars are listed in Table\,\ref{table:Mnpec}. The dash-dot curve is a rough envelope of observed overabundances. Notice the small deep that could be guessed around $T_{\rm{eff}}=13100$\,K that motivates our speculation (see text).}
\label{fig:Mnpec2}
\end{figure}

Manganese is an interesting element since its systematic abundance peculiarities shown in Fig.\,\ref{fig:Mnpec2} vs. $T_{\rm{eff}}$, characterize (with Hg) the HgMn CP-type. It is also among the first elements for which a quantitative theoretical prediction was successfully done in the framework of atomic diffusion processes \citep{AlecianAlMi1981}. These authors computed the maximum overabundance of Mn that can be supported by the radiation field in the atmospheres of HgMn stars (dashed curve in Fig.\,\ref{fig:Mnpec2}). This prediction was confirmed through spectroscopic measurements by \citet{SmithSmDw1993}.

\citet{SmithSmDw1993} also noticed in the range 13000K$\,\lta T_{\mathrm{eff}} \lta$\,15000K some stars in which the overabundance of Mn is significantly smaller than in the others. They suggested, following \citet{CowleyCo1980}, that these stars may form {\it{a distinct group of hot, mild Mn stars}}. These stars are highlighted in Fig.\,\ref{fig:Mnpec2} and we are inclined to identify a first group A (purple circles surrounding crosses) and a second group B (red squares  surrounding crosses), both groups are listed in Table\,\ref{table:Mnpec}. We justify our choice of defining these two groups by the following reasons. Stars of group A are found very close to $T_{\rm{eff}}=13100K$ and well gathered, their $\log{g}$ cover almost the whole range of gravity found for HgMn stars (from $3.6$ to $4.29$), and with various $v\,\sin{i}$ from very small to very large (for HgMn stars). The group B consists in the 3 hottest highlighted stars around $T_{\rm{eff}}=14000$\,K with $\log{g}$ or $v\,\sin{i}$ close to extrema. We make the hypothesis that group B is not related to group A. We also make the following speculation based on the rough envelope of observed overabundances shown by dash-dot curve in Fig.\,\ref{fig:Mnpec2}: we notice that considering this envelope, the Mn overabundances seem to form around $T_{\rm{eff}}=13100$\,K a small deep even for stars that do not show Mn overabundance mildness (they are not included in group A). Therefore, we suggest that, rather to consider the existence of {\it{a distinct group of hot, mild Mn stars}} (actually our group A), there is a unique population of HgMn stars, but some physical process happens in atmospheres around $T_{\rm{eff}}=13100$\,K that causes depressed Mn overabundances around this effective temperature. Such a situation is compatible with atomic diffusion theory, which is suspected to form element stratifications potentially unstable in atmospheres \citep{AlecianGeStMaDorfiEA2011}. Of course, we cannot exclude that this gathering around this particular $T_{\rm{eff}}$, and definition of group A and B, might be an artifact of effective temperature determination methods and related errors.

\begin{table}
\begin{center}
\caption{List of highlighted stars in Fig.\,\ref{fig:Mnpec2}. Stars of group A are very close to $T_{\rm{eff}}=13100$\,K, the 3 hottest highlighted stars form the group B (see text).}
\label{table:Mnpec} 
\begin{tabular}{|l|c|c|c|c|c|c|}
\hline
Star & group & $T_{\rm{eff}}$ & $\log{g}$ & $v\,\sin{i}$ & Mult. & $\epsilon$(Mn)\\  \hline
HD\,46886 & A & 12900 & 3.8 & 18 & 1 & 1.94\\
HD\,186122 & A & 13000 & 3.65 & 3 & 1 & 0.77\\
HD\,55362 & A & 13000 & 4 & 53 & 1 & 0.81\\
HD\,144218 & A & 13000 & 4.2 & 5 & 5 & 1.7\\
HD\,196426 & A & 13010 & 3.84 & 5.5 & 1 & 0.25\\
HD\,174933\,A & A & 13100 & 4.1 & 6 & 4 & 1.74\\
HD\,7374 & A & 13150 & 4 & 21 & 4 & 1.87\\
HD\,179761 & A & 13175 & 3.27 & 17 & 2 & 0.83\\
HD\,190229\,B & A & 13200 & 3.6 & 8 & 4 & 1.15\\
HD\,35497 & A & 13250 & 3.65 & 59 & 2 & 1.73\\
HD\,37492\,A & A & 13325 & 3.93 & 8.5 & 3 & 1.2\\
HD\,79158 & B & 13700 & 3.65 & 49 & 1 & -0.23\\
HD\,144667 & B & 13990 & 4.29 & 1.5 & 5 & 1.27\\
HD\,53929 & B & 14050 & 3.6 & 21 & 1 & 0.73\\
\hline
\end{tabular}
\end{center}
\end{table}

\subsection{The case of calcium in AmFm stars}
\label{Cacase}

\begin{figure}
\includegraphics[width=9cm]{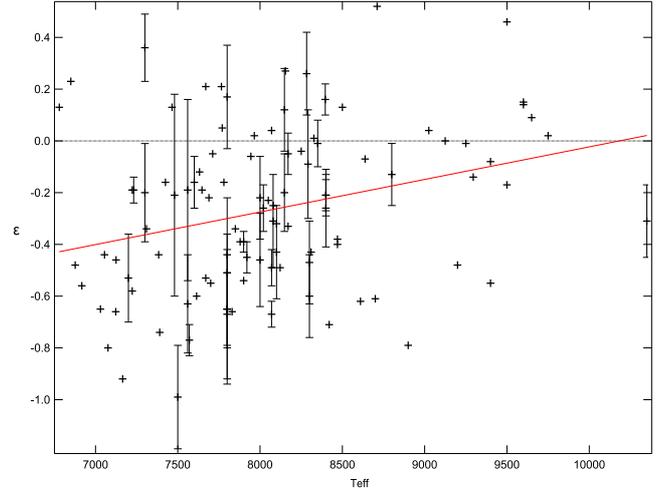}
\caption{Ca abundances ($\epsilon$) in AmFm stars vs. $T_{\rm{eff}}$. The red solid line is a linear fit showing the correlation of $\epsilon$ and $T_{\rm{eff}}$.}
\label{fig:Ca_AmFm}
\end{figure}

Calcium shown in Fig.\,\ref{fig:Ca_AmFm} is an important element for AmFm stars (as Mn for HgMn), since it helps in determining the AmFm group due to its systematic underabundances. According to models including atomic diffusion, this underabundance is caused by the small radiative acceleration on Ca just below the superficial convection zone where Ca is mainly in ionization stage having noble-gas configuration \citep[see for instance][]{AlecianAl1996}. Considering increasing effective temperatures, the bottom of the superficial convection zone moves to higher stellar layers where Ca ionization stages are no more dominated by stages in noble-gas configuration. Then, the underabundance of Ca is reduced until the disappearance of the AmFm phenomenon (above $T_{\rm{eff}}\approx 10000$\,K) and the appearance of the HgMn phenomenon.  Fig.\,\ref{fig:Ca_AmFm} is possibly showing this transition between AmFm and HgMn CP-types.

\subsection{The case of helium in ApBp stars}
\label{Hecase}

\begin{figure}
\includegraphics[width=9cm]{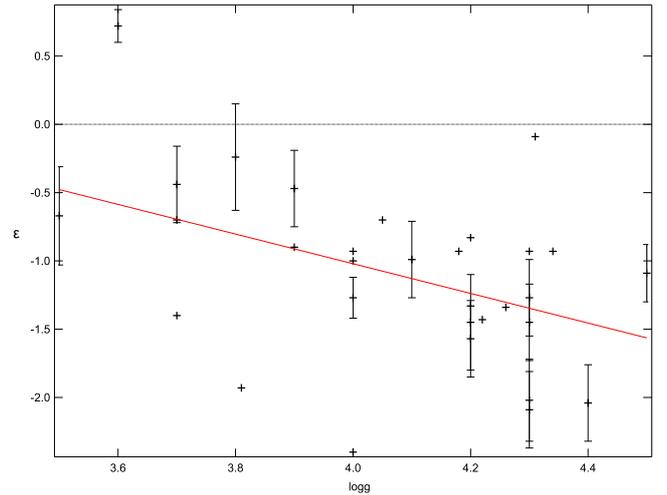}
\caption{He abundances ($\epsilon$) in ApBp stars vs. $\log{g}$. The red solid line is a linear fit showing the correlation of $\epsilon$ and gravity.}
\label{fig:He_ApBp}
\end{figure}

Helium plays an important role in the scenario leading to the CP phenomenon in the framework of atomic diffusion. Because, CP stars are slow rotators, and then rotational mixing is weaker than in non-CP stars, and because helium is not supported by the radiation field in upper layers (small radiative acceleration), it undergoes essentially gravitational settling. This leads to a systematic underabundance of He (see Fig.~\ref{fig:Layout_panels}) in all the three CP-type stars we consider. Such a He deficiency reduces significantly the superficial convection zones, which helps in stabilizing external layers and so, helps atomic diffusion to be efficient \citep[see][for a complete discussion]{MichaudMiAlRi2015}.

In  Fig.~\ref{fig:He_ApBp} one can notice the clear trend of He abundance to decrease in ApBp stars with increasing gravity (increasing efficiency of gravitational settling). Such a trend is not visible for AmFm and HgMn stars. This may be explained by the stabilizing mechanism of upper layers that are different in ApBp stars than in the other CP-type stars: in ApBp stars magnetic fields are usually considered to be strongly involved in stabilizing external layers.

\subsection{The case of iron in ApBp stars}
\label{Fecase}

\begin{figure}
\includegraphics[width=9cm]{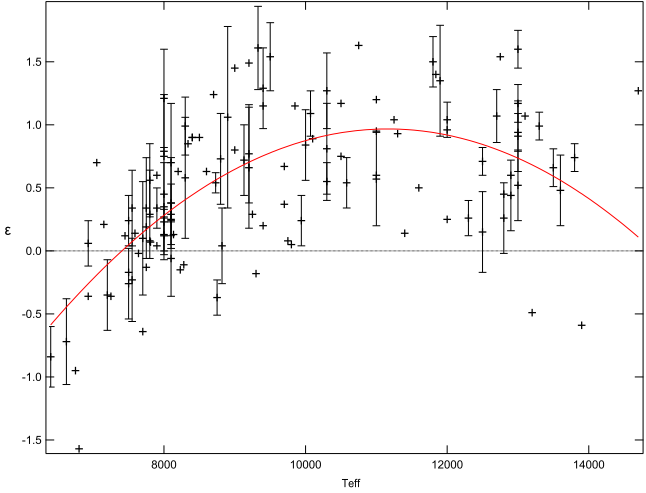}
\caption{Fe abundances ($\epsilon$) in ApBp stars vs. $T_{\rm{eff}}$. The red solid curve is a polynomial fit showing the correlation of $\epsilon$ and $T_{\rm{eff}}$.}
\label{fig:Fe_ApBp}
\end{figure}

Radiative acceleration for iron is not often as strong as for other iron peak metals, despite of a very large number of bound-bound atomic transitions contributing to transfer momentum from photons to iron atoms. This is due to the large cosmic abundance of Fe in stars having solar metallicity (radiative acceleration decreases when the number of absorbing atoms increases), at least when they arrive on the main-sequence. However, Fe is generally enough supported by the radiation fields, and that leads to observe Fe overabundances in CP stars. But some underabundances are also observed. The coexistence of both over and underabundances reflects the complexity of the abundance stratifications build-up in atmospheres, especially when an element is marginally supported by the radiation field. 

In  Fig.~\ref{fig:Fe_ApBp}, we show Fe abundances ($\epsilon$) in ApBp stars plotted vs. $T_{\rm{eff}}$. We have determined a 2nd order polynomial fit (solid red curve) to emphasize the global trend we think to observe with respect to the effective temperature (not for the other parameters). This trend shows a clear maximum around $T_{\rm{eff}}=11000$\,K. Such maximum is not found for AmFm and HgMn stars, that leads to suspect again the magnetic fields to cause such a behavior, since the diffusion velocity is strongly affected by the magnetic field in the upper photosphere \citep[see for instance][for a recent discussion on the subject]{AlecianAl2015}.

\section{Conclusions}
\label{sec:conc}

This work presents a unique catalogue of 428 Chemically Peculiar stars observed by spectroscopy during the last decades. We have compiled the main physical parameters and the abundances of elements from helium to uranium for 108 HgMn, 188 ApBp, 120 AmFm stars, plus 12 other peculiar stars (including 1 horizontal-branch star). This new compilation confirms (as in our Paper I devoted to HgMn stars) the increase of overabundances for heavy elements with atomic number \citep[shown by][]{SmithSm1996a} and the large scatter of the abundances anomalies. This scatter is not only due to the heterogeneity of the data, or abundance determination errors, but real. 

We have applied AD test on single CP stars and those being in binary systems and do not find any relation between abundance anomalies and multiplicity in CP1, CP2, and CP3 type stars, possibly because of the lack of data.

To check if abundance anomalies are correlated with fundamental parameters, we applied the Spearman's rank correlation test between abundances and physical parameters, such as effective temperature, surface gravity and rotation velocity. We have found a significant number of correlations with the effective temperatures, but also some (fewer) with gravity and rotational velocity. According to usual models \citep{MichaudMiAlRi2015}, correlations of abundances with $v\,\sin{i}$ could be rather related to rotational mixing, those with $\log{g}$ to the stellar structure an the competition with radiative accelerations,  and correlations with $T_{\rm{eff}}$ to radiative accelerations. Of course, in a given star, all of these parameters should have to be combined. We have also chosen to discuss and comment  four noteworthy cases, but this does't mean that there are only four cases that deserve discussion. We are convinced that considering as a whole, the abundance measurements in CP stars will lead to interesting understanding of the physical processes in play in these stars. In the near future we intend to extend our database to new observations and to other categories of chemically peculiar stars (such as stars with helium anomalies), and we have no doubt that a major extension of such a database will be achieved from the final GAIA catalogue.

\section*{Acknowledgements}

This work was supported by the RA MES State Committee of Science, in the frames of the research project N\textsuperscript{\underline{o}} 16YR-1C034.

%%%%%%%%%%%%%%%%%%%%%%%%%%%%%%%%%%%%%%%%%%%%%%%%%%


\bibliographystyle{mnras}
\bibliography{cp_stars}

\begin{thebibliography}{}
\makeatletter
\relax
\def\mn@urlcharsother{\let\do\@makeother \do\$\do\&\do\#\do\^\do\_\do\%\do\~}
\def\mn@doi{\begingroup\mn@urlcharsother \@ifnextchar [ {\mn@doi@}
  {\mn@doi@[]}}
\def\mn@doi@[#1]#2{\def\@tempa{#1}\ifx\@tempa\@empty \href
  {http://dx.doi.org/#2} {doi:#2}\else \href {http://dx.doi.org/#2} {#1}\fi
  \endgroup}
\def\mn@eprint#1#2{\mn@eprint@#1:#2::\@nil}
\def\mn@eprint@arXiv#1{\href {http://arxiv.org/abs/#1} {{\tt arXiv:#1}}}
\def\mn@eprint@dblp#1{\href {http://dblp.uni-trier.de/rec/bibtex/#1.xml}
  {dblp:#1}}
\def\mn@eprint@#1:#2:#3:#4\@nil{\def\@tempa {#1}\def\@tempb {#2}\def\@tempc
  {#3}\ifx \@tempc \@empty \let \@tempc \@tempb \let \@tempb \@tempa \fi \ifx
  \@tempb \@empty \def\@tempb {arXiv}\fi \@ifundefined
  {mn@eprint@\@tempb}{\@tempb:\@tempc}{\expandafter \expandafter \csname
  mn@eprint@\@tempb\endcsname \expandafter{\@tempc}}}

\bibitem[\protect\citeauthoryear{{Adelman}, {Philip}  \& {Adelman}}{{Adelman}
  et~al.}{1996}]{AdelmanAdPhAd1996}
{Adelman} S.~J.,  {Philip} A.~G.~D.,   {Adelman} C.~J.,  1996, \mn@doi [\mnras]
  {10.1093/mnras/282.3.953}, \href
  {http://adsabs.harvard.edu/abs/1996MNRAS.282..953A} {282, 953}

\bibitem[\protect\citeauthoryear{{Adelman}, {Caliskan}, {Cay}, {Kocer}  \&
  {Tektanali}}{{Adelman} et~al.}{1999}]{AdelmanAdCaCaetal1999}
{Adelman} S.~J.,  {Caliskan} H.,  {Cay} T.,  {Kocer} D.,   {Tektanali} H.~G.,
  1999, \mn@doi [\mnras] {10.1046/j.1365-8711.1999.02435.x}, \href
  {http://adsabs.harvard.edu/abs/1999MNRAS.305..591A} {305, 591}

\bibitem[\protect\citeauthoryear{{Alecian}}{{Alecian}}{1996}]{AlecianAl1996}
{Alecian} G.,  1996, A\&A, \href
  {http://adsabs.harvard.edu/abs/1996A%26A...310..872A} {310, 872}

\bibitem[\protect\citeauthoryear{{Alecian}}{{Alecian}}{2015}]{AlecianAl2015}
{Alecian} G.,  2015, \mn@doi [MNRAS] {10.1093/mnras/stv2205}, \href
  {http://adsabs.harvard.edu/abs/2015MNRAS.454.3143A} {454, 3143}

\bibitem[\protect\citeauthoryear{{Alecian} \& {Michaud}}{{Alecian} \&
  {Michaud}}{1981}]{AlecianAlMi1981}
{Alecian} G.,  {Michaud} G.,  1981, \mn@doi [ApJ] {10.1086/158803}, \href
  {http://adsabs.harvard.edu/abs/1981ApJ...245..226A} {245, 226}

\bibitem[\protect\citeauthoryear{{Alecian}, {Gebran}, {Auvergne}, {Richard},
  {Samadi}, {Weiss}  \& {Baglin}}{{Alecian}
  et~al.}{2009}]{AlecianAlGeAuetal2009}
{Alecian} G.,  {Gebran} M.,  {Auvergne} M.,  {Richard} O.,  {Samadi} R.,
  {Weiss} W.~W.,   {Baglin} A.,  2009, \mn@doi [\aap]
  {10.1051/0004-6361/200911678}, \href
  {http://adsabs.harvard.edu/abs/2009A%26A...506...69A} {506, 69}

\bibitem[\protect\citeauthoryear{{Alecian}, {Stift}  \& {Dorfi}}{{Alecian}
  et~al.}{2011}]{AlecianGeStMaDorfiEA2011}
{Alecian} G.,  {Stift} M.~J.,   {Dorfi} E.~A.,  2011, \mn@doi [\mnras]
  {10.1111/j.1365-2966.2011.19547.x}, \href
  {http://adsabs.harvard.edu/abs/2011MNRAS.418..986A} {418, 986}

\bibitem[\protect\citeauthoryear{{Alecian}, {Tkachenko}, {Neiner}, {Folsom}  \&
  {Leroy}}{{Alecian} et~al.}{2016}]{AlecianAlTkNeetal2016}
{Alecian} E.,  {Tkachenko} A.,  {Neiner} C.,  {Folsom} C.~P.,   {Leroy} B.,
  2016, \mn@doi [\aap] {10.1051/0004-6361/201527355}, \href
  {http://adsabs.harvard.edu/abs/2016A%26A...589A..47A} {589, A47}

\bibitem[\protect\citeauthoryear{{Alentiev}, {Kochukhov}, {Ryabchikova},
  {Cunha}, {Tsymbal}  \& {Weiss}}{{Alentiev}
  et~al.}{2012}]{AlentievAlKoRyetal2012}
{Alentiev} D.,  {Kochukhov} O.,  {Ryabchikova} T.,  {Cunha} M.,  {Tsymbal} V.,
   {Weiss} W.,  2012, \mn@doi [\mnras] {10.1111/j.1745-3933.2011.01211.x},
  \href {http://adsabs.harvard.edu/abs/2012MNRAS.421L..82A} {421, L82}

\bibitem[\protect\citeauthoryear{{Asplund}, {Grevesse}, {Sauval}  \&
  {Scott}}{{Asplund} et~al.}{2009}]{AsplundMaGrSaSc2009}
{Asplund} M.,  {Grevesse} N.,  {Sauval} A.~J.,   {Scott} P.,  2009, \mn@doi
  [\araa] {10.1146/annurev.astro.46.060407.145222}, \href
  {http://adsabs.harvard.edu/abs/2009ARA%26A..47..481A} {47, 481}

\bibitem[\protect\citeauthoryear{{Bolcal}, {Kocer}  \& {Duzgelen}}{{Bolcal}
  et~al.}{1987}]{BolcalBoKoDu1987}
{Bolcal} C.,  {Kocer} D.,   {Duzgelen} A.,  1987, \mn@doi [\apss]
  {10.1007/BF00644358}, \href
  {http://adsabs.harvard.edu/abs/1987Ap%26SS.139..295B} {139, 295}

\bibitem[\protect\citeauthoryear{{Borsenberger}, {Michaud}  \&
  {Praderie}}{{Borsenberger} et~al.}{1979}]{BorsenbergerBoMiPr1979}
{Borsenberger} J.,  {Michaud} G.,   {Praderie} F.,  1979, \aap, \href
  {http://adsabs.harvard.edu/abs/1979A%26A....76..287B} {76, 287}

\bibitem[\protect\citeauthoryear{{Budaj} \& {Iliev}}{{Budaj} \&
  {Iliev}}{2003}]{BudajBuIl2003}
{Budaj} J.,  {Iliev} I.~K.,  2003, \mn@doi [\mnras]
  {10.1046/j.1365-2966.2003.07071.x}, \href
  {http://adsabs.harvard.edu/abs/2003MNRAS.346...27B} {346, 27}

\bibitem[\protect\citeauthoryear{{Burkhart} \& {Coupry}}{{Burkhart} \&
  {Coupry}}{1989}]{BurkhartBuCo1989}
{Burkhart} C.,  {Coupry} M.~F.,  1989, \aap, \href
  {http://adsabs.harvard.edu/abs/1989A%26A...220..197B} {220, 197}

\bibitem[\protect\citeauthoryear{{Castelli} \& {Hubrig}}{{Castelli} \&
  {Hubrig}}{2004}]{CastelliCaHu2004}
{Castelli} F.,  {Hubrig} S.,  2004, \mn@doi [\aap]
  {10.1051/0004-6361:20041011}, \href
  {http://adsabs.harvard.edu/abs/2004A%26A...425..263C} {425, 263}

\bibitem[\protect\citeauthoryear{{Catanzaro}, {Leone}  \& {Leto}}{{Catanzaro}
  et~al.}{2003}]{CatanzaroCaLeLe2003}
{Catanzaro} G.,  {Leone} F.,   {Leto} P.,  2003, \mn@doi [\aap]
  {10.1051/0004-6361:20030887}, \href
  {http://adsabs.harvard.edu/abs/2003A%26A...407..669C} {407, 669}

\bibitem[\protect\citeauthoryear{{Catanzaro}, {Giarrusso}, {Leone}, {Munari},
  {Scalia}, {Sparacello}  \& {Scuderi}}{{Catanzaro}
  et~al.}{2016}]{CatanzaroCaGiLeetal2016}
{Catanzaro} G.,  {Giarrusso} M.,  {Leone} F.,  {Munari} M.,  {Scalia} C.,
  {Sparacello} E.,   {Scuderi} S.,  2016, \mn@doi [\mnras]
  {10.1093/mnras/stw923}, \href
  {http://adsabs.harvard.edu/abs/2016MNRAS.460.1999C} {460, 1999}

\bibitem[\protect\citeauthoryear{{Coupry}, {vant Veer-Menneret}  \&
  {Burkhart}}{{Coupry} et~al.}{1986}]{CoupryCovaBu1986}
{Coupry} M.~F.,  {vant Veer-Menneret} C.,   {Burkhart} C.,  1986, \aaps, \href
  {http://adsabs.harvard.edu/abs/1986A%26AS...64..477C} {64, 477}

\bibitem[\protect\citeauthoryear{{Cowley}}{{Cowley}}{1980}]{CowleyCo1980}
{Cowley} C.~R.,  1980, \mn@doi [\pasp] {10.1086/130640}, \href
  {http://adsabs.harvard.edu/abs/1980PASP...92..159C} {92, 159}

\bibitem[\protect\citeauthoryear{{Dolk}, {Wahlgren}  \& {Hubrig}}{{Dolk}
  et~al.}{2003}]{DolkDoWaHu2003}
{Dolk} L.,  {Wahlgren} G.~M.,   {Hubrig} S.,  2003, \mn@doi [\aap]
  {10.1051/0004-6361:20030213}, \href
  {http://adsabs.harvard.edu/abs/2003A%26A...402..299D} {402, 299}

\bibitem[\protect\citeauthoryear{{Dworetsky}, {Dyer}  \& {Persaud}}{{Dworetsky}
  et~al.}{2008a}]{DworetskyDwDyPe2008}
{Dworetsky} M.~M.,  {Dyer} A.,   {Persaud} J.~L.,  2008a, Contributions of the
  Astronomical Observatory Skalnate Pleso, \href
  {http://adsabs.harvard.edu/abs/2008CoSka..38..141D} {38, 141}

\bibitem[\protect\citeauthoryear{{Dworetsky}, {Persaud}  \&
  {Patel}}{{Dworetsky} et~al.}{2008b}]{DworetskyDwPePa2008}
{Dworetsky} M.~M.,  {Persaud} J.~L.,   {Patel} K.,  2008b, \mn@doi [\mnras]
  {10.1111/j.1365-2966.2008.12937.x}, \href
  {http://adsabs.harvard.edu/abs/2008MNRAS.385.1523D} {385, 1523}

\bibitem[\protect\citeauthoryear{{Engmann} \& {Cousineau}}{{Engmann} \&
  {Cousineau}}{2011}]{EngmannCousineau11}
{Engmann} S.,  {Cousineau} D.,  2011, J. Appl. Quant. Methods, 6, 1

\bibitem[\protect\citeauthoryear{{Feigelson} \& {Babu}}{{Feigelson} \&
  {Babu}}{2012}]{FeigelsonBabu2012}
{Feigelson} E.~D.,  {Babu} G.~J.,  2012, {Modern Statistical Methods for
  Astronomy, Cambridge, UK: Cambridge University Press}

\bibitem[\protect\citeauthoryear{{Folsom} et~al.,}{{Folsom}
  et~al.}{2007}]{FolsomFoWaHaetal2007}
{Folsom} C.~F.,  et~al., 2007, in {Romanyuk} I.~I.,  {Kudryavtsev} D.~O.,
  {Neizvestnaya} O.~M.,   {Shapoval} V.~M.,  eds, Physics of Magnetic Stars. pp
  189--196 (\mn@eprint {} {astro-ph/0612231})

\bibitem[\protect\citeauthoryear{{Gebran}, {Monier}  \& {Richard}}{{Gebran}
  et~al.}{2008}]{GebranGeMoRi2008}
{Gebran} M.,  {Monier} R.,   {Richard} O.,  2008, \mn@doi [\aap]
  {10.1051/0004-6361:20078807}, \href
  {http://adsabs.harvard.edu/abs/2008A%26A...479..189G} {479, 189}

\bibitem[\protect\citeauthoryear{{Gelbmann}}{{Gelbmann}}{1998}]{GelbmannGe1998}
{Gelbmann} M.~J.,  1998, Contributions of the Astronomical Observatory Skalnate
  Pleso, \href {http://adsabs.harvard.edu/abs/1998CoSka..27..280G} {27, 280}

\bibitem[\protect\citeauthoryear{{Ghazaryan} \& {Alecian}}{{Ghazaryan} \&
  {Alecian}}{2016}]{GhazaryanGhAl2016}
{Ghazaryan} S.,  {Alecian} G.,  2016, \mn@doi [\mnras] {10.1093/mnras/stw911},
  \href {http://adsabs.harvard.edu/abs/2016MNRAS.460.1912G} {460, 1912}

\bibitem[\protect\citeauthoryear{{Hubrig}, {Castelli}  \& {Mathys}}{{Hubrig}
  et~al.}{1999}]{HubrigHuCaMa1999}
{Hubrig} S.,  {Castelli} F.,   {Mathys} G.,  1999, \aap, \href
  {http://adsabs.harvard.edu/abs/1999A%26A...341..190H} {341, 190}

\bibitem[\protect\citeauthoryear{{Hubrig}, {Castelli}, {Gonz{\'a}lez}, {Elkin},
  {Mathys}, {Cowley}, {Wolff}  \& {Sch{\"o}ller}}{{Hubrig}
  et~al.}{2012}]{HubrigHuCaGoetal2012}
{Hubrig} S.,  {Castelli} F.,  {Gonz{\'a}lez} J.~F.,  {Elkin} V.~G.,  {Mathys}
  G.,  {Cowley} C.~R.,  {Wolff} B.,   {Sch{\"o}ller} M.,  2012, \mn@doi [\aap]
  {10.1051/0004-6361/201218968}, \href
  {http://adsabs.harvard.edu/abs/2012A%26A...542A..31H} {542, A31}

\bibitem[\protect\citeauthoryear{{Kochukhov}, {Tsymbal}, {Ryabchikova},
  {Makaganyk}  \& {Bagnulo}}{{Kochukhov}
  et~al.}{2006}]{KochukhovKoTsRyetal2006}
{Kochukhov} O.,  {Tsymbal} V.,  {Ryabchikova} T.,  {Makaganyk} V.,   {Bagnulo}
  S.,  2006, \mn@doi [\aap] {10.1051/0004-6361:20065607}, \href
  {http://adsabs.harvard.edu/abs/2006A%26A...460..831K} {460, 831}

\bibitem[\protect\citeauthoryear{{Kunzli} \& {North}}{{Kunzli} \&
  {North}}{1998}]{KunzliKuNo1998}
{Kunzli} M.,  {North} P.,  1998, \aap, \href
  {http://adsabs.harvard.edu/abs/1998A%26A...330..651K} {330, 651}

\bibitem[\protect\citeauthoryear{{Kupka}, {Ryabchikova}, {Bolgova}, {Kuschnig},
  {Weiss}, {Mathys}  \& {Le Contel}}{{Kupka}
  et~al.}{1994}]{KupkaKuRyBoetal1994}
{Kupka} F.,  {Ryabchikova} T.,  {Bolgova} G.,  {Kuschnig} R.,  {Weiss} W.~W.,
  {Mathys} G.,   {Le Contel} J.~M.,  1994, in {Zverko} J.,  {Ziznovsky} J.,
  eds, Chemically Peculiar and Magnetic Stars. p.~130

\bibitem[\protect\citeauthoryear{{LeBlanc}, {Khalack}, {Yameogo}, {Thibeault}
  \& {Gallant}}{{LeBlanc} et~al.}{2015}]{LeBlancLeKhYaetal2015}
{LeBlanc} F.,  {Khalack} V.,  {Yameogo} B.,  {Thibeault} C.,   {Gallant} I.,
  2015, \mn@doi [\mnras] {10.1093/mnras/stv1466}, \href
  {http://adsabs.harvard.edu/abs/2015MNRAS.453.3766L} {453, 3766}

\bibitem[\protect\citeauthoryear{{Leckrone}}{{Leckrone}}{1981}]{LeckroneLe1981}
{Leckrone} D.~S.,  1981, \mn@doi [\apj] {10.1086/159415}, \href
  {http://adsabs.harvard.edu/abs/1981ApJ...250..687L} {250, 687}

\bibitem[\protect\citeauthoryear{{Makaganiuk} et~al.,}{{Makaganiuk}
  et~al.}{2012}]{MakaganiukMaKoPietal2012}
{Makaganiuk} V.,  et~al., 2012, \mn@doi [\aap] {10.1051/0004-6361/201118167},
  \href {http://adsabs.harvard.edu/abs/2012A%26A...539A.142M} {539, A142}

\bibitem[\protect\citeauthoryear{{Michaud}, {Alecian}  \& {Richer}}{{Michaud}
  et~al.}{2015}]{MichaudMiAlRi2015}
{Michaud} G.,  {Alecian} G.,   {Richer} J.,  2015, {Atomic Diffusion in Stars,
  Astronomy and Astrophysics Library, Springer International Publishing,
  Switzerland.}, \mn@doi{10.1007/978-3-319-19854-5.
}

\bibitem[\protect\citeauthoryear{{Monier} \& {Megessier}}{{Monier} \&
  {Megessier}}{1990}]{MonierMoMe1990}
{Monier} R.,  {Megessier} C.,  1990, \aap, \href
  {http://adsabs.harvard.edu/abs/1990A%26A...237..402M} {237, 402}

\bibitem[\protect\citeauthoryear{{Ndiaye}, {LeBlanc}  \& {Khalack}}{{Ndiaye}
  et~al.}{2018}]{NdiayeNdLeKh2018}
{Ndiaye} M.~L.,  {LeBlanc} F.,   {Khalack} V.,  2018, \mn@doi [\mnras]
  {10.1093/mnras/sty693}, \href
  {http://adsabs.harvard.edu/abs/2018MNRAS.477.3390N} {477, 3390}

\bibitem[\protect\citeauthoryear{{Niemczura}, {Morel}  \& {Aerts}}{{Niemczura}
  et~al.}{2009}]{NiemczuraNiMoAe2009}
{Niemczura} E.,  {Morel} T.,   {Aerts} C.,  2009, \mn@doi [\aap]
  {10.1051/0004-6361/200911931}, \href
  {http://adsabs.harvard.edu/abs/2009A%26A...506..213N} {506, 213}

\bibitem[\protect\citeauthoryear{{Niemczura} et~al.,}{{Niemczura}
  et~al.}{2015}]{NiemczuraNiMuSmetal2015}
{Niemczura} E.,  et~al., 2015, \mn@doi [\mnras] {10.1093/mnras/stv528}, \href
  {http://adsabs.harvard.edu/abs/2015MNRAS.450.2764N} {450, 2764}

\bibitem[\protect\citeauthoryear{{Nishimura}, {Sadakane}, {Kato}, {Takeda}  \&
  {Mathys}}{{Nishimura} et~al.}{2004}]{NishimuraNiSaKaetal2004}
{Nishimura} M.,  {Sadakane} K.,  {Kato} K.,  {Takeda} Y.,   {Mathys} G.,  2004,
  \mn@doi [\aap] {10.1051/0004-6361:20035651}, \href
  {http://adsabs.harvard.edu/abs/2004A%26A...420..673N} {420, 673}

\bibitem[\protect\citeauthoryear{{Polosukhina} et~al.,}{{Polosukhina}
  et~al.}{2004}]{PolosukhinaPoShDretal2004}
{Polosukhina} N.,  et~al., 2004, in {Zverko} J.,  {Ziznovsky} J.,  {Adelman}
  S.~J.,   {Weiss} W.~W.,  eds,  IAU Symposium Vol. 224, The A-Star Puzzle. pp
  665--672, \mn@doi{10.1017/S1743921305009555}

\bibitem[\protect\citeauthoryear{{Preston}}{{Preston}}{1974}]{1974Preston}
{Preston} G.~W.,  1974, \mn@doi [\araa] {10.1146/annurev.aa.12.090174.001353},
  \href {http://adsabs.harvard.edu/abs/1974ARA%26A..12..257P} {12, 257}

\bibitem[\protect\citeauthoryear{{Proffitt} \& {Michaud}}{{Proffitt} \&
  {Michaud}}{1991}]{ProffittPrMi1991}
{Proffitt} C.~R.,  {Michaud} G.,  1991, \mn@doi [\apj] {10.1086/169923}, \href
  {http://adsabs.harvard.edu/abs/1991ApJ...371..584P} {371, 584}

\bibitem[\protect\citeauthoryear{{Renson} \& {Manfroid}}{{Renson} \&
  {Manfroid}}{2009}]{2009RensonandMJ}
{Renson} P.,  {Manfroid} J.,  2009, \mn@doi [A\&A]
  {10.1051/0004-6361/200810788}, 498, 961

\bibitem[\protect\citeauthoryear{{Ryabchikova}}{{Ryabchikova}}{1998}]{RyabchikovaRy1998}
{Ryabchikova} T.,  1998, Contributions of the Astronomical Observatory Skalnate
  Pleso, \href {http://adsabs.harvard.edu/abs/1998CoSka..27..319R} {27, 319}

\bibitem[\protect\citeauthoryear{{Ryabchikova}, {Piskunov}, {Savanov}, {Kupka}
  \& {Malanushenko}}{{Ryabchikova} et~al.}{1999}]{RyabchikovaRyPiSaetal1999}
{Ryabchikova} T.,  {Piskunov} N.,  {Savanov} I.,  {Kupka} F.,   {Malanushenko}
  V.,  1999, \aap, \href {http://adsabs.harvard.edu/abs/1999A%26A...343..229R}
  {343, 229}

\bibitem[\protect\citeauthoryear{{Smith}}{{Smith}}{1993}]{SmithSm1993}
{Smith} K.~C.,  1993, \aap, \href
  {http://adsabs.harvard.edu/abs/1993A%26A...276..393S} {276, 393}

\bibitem[\protect\citeauthoryear{{Smith}}{{Smith}}{1996a}]{SmithSm1996a}
{Smith} K.~C.,  1996a, \mn@doi [\apss] {10.1007/BF02424427}, \href
  {http://adsabs.harvard.edu/abs/1996Ap%26SS.237...77S} {237, 77}

\bibitem[\protect\citeauthoryear{{Smith}}{{Smith}}{1996b}]{SmithSm1996}
{Smith} K.~C.,  1996b, \aap, \href
  {http://adsabs.harvard.edu/abs/1996A%26A...305..902S} {305, 902}

\bibitem[\protect\citeauthoryear{{Smith} \& {Dworetsky}}{{Smith} \&
  {Dworetsky}}{1993}]{SmithSmDw1993}
{Smith} K.~C.,  {Dworetsky} M.~M.,  1993, \aap, \href
  {http://adsabs.harvard.edu/abs/1993A%26A...274..335S} {274, 335}

\bibitem[\protect\citeauthoryear{{Spearman}}{{Spearman}}{1904}]{Spearman1904}
{Spearman} C.,  1904, American Journal of Psychology, 15, 72

\bibitem[\protect\citeauthoryear{{Spite} \& {Spite}}{{Spite} \&
  {Spite}}{1982}]{SpiteSpSp1982}
{Spite} F.,  {Spite} M.,  1982, \aap, \href
  {http://adsabs.harvard.edu/abs/1982A%26A...115..357S} {115, 357}

\bibitem[\protect\citeauthoryear{{Y{\"u}ce}, {Castelli}  \&
  {Hubrig}}{{Y{\"u}ce} et~al.}{2011}]{YuceYuCaHu2011}
{Y{\"u}ce} K.,  {Castelli} F.,   {Hubrig} S.,  2011, \mn@doi [\aap]
  {10.1051/0004-6361/201016251}, \href
  {http://adsabs.harvard.edu/abs/2011A%26A...528A..37Y} {528, A37}

\makeatother
\end{thebibliography}

%\cite{*}
%\end{thebibliography}

%%%%%%%%%%%%%%%%%%%%%%%%%%%%%%%%%%%%%%%%%%%%%%%%%%

%%%%%%%%%%%%%%%%% APPENDICES %%%%%%%%%%%%%%%%%%%%%

%\appendix

%\section{}
%References:


%%%%%%%%%%%%%%%%%%%%%%%%%%%%%%%%%%%%%%%%%%%%%%%%%%

% Don't change these lines
\bsp	% typesetting comment
\label{lastpage}
\end{document}